\documentclass[aps,prb,twocolumn,superscriptaddress,amsmath,amssymb,showpacs]{revtex4-1}
\usepackage[english]{babel}
\usepackage{amssymb}
\usepackage{amsmath}
\usepackage[dvips]{graphicx}
\usepackage{dcolumn}
\usepackage{bm}
\usepackage{geometry}

\begin{document}

\title{Model independent X-ray standing wave analysis of periodic multilayer structures}

\author{S.N.~Yakunin}	
\affiliation{NRC Kurchatov Institute, Moscow, Russia}

\author{I.A.~Makhotkin}	
\affiliation{FOM Institute DIFFER, Nieuwegein, Netherlands}

\author{M.A.~Chuev}	
\affiliation{IPT RAS, Moscow, Russia}

\author{E.M.~Pashaev}	
\affiliation{NRC Kurchatov Institute, Moscow, Russia}

\author{E.~Zoethout}	
\affiliation{FOM Institute DIFFER, Nieuwegein, Netherlands}

\author{E.~Louis}	
\affiliation{FOM Institute DIFFER, Nieuwegein, Netherlands}
\affiliation{MESA+ Institute for Nanotechnology, University of Twente, Enschede, Netherlands}

\author{R.W.E.~van~de~Kruijs}	
\affiliation{FOM Institute DIFFER, Nieuwegein, Netherlands}

\author{S.Yu.~Seregin}	
\affiliation{IC RAS, Moscow, Russia}

\author{I.A.~Subbotin}	
\affiliation{NRC Kurchatov Institute, Moscow, Russia}

\author{D.V.~Novikov}	
\affiliation{DESY, Hamburg, Germany}

\author{F.~Bijkerk}	
\affiliation{FOM Institute DIFFER, Nieuwegein, Netherlands}
\affiliation{MESA+ Institute for Nanotechnology, University of Twente, Enschede, Netherlands}

\author{M.V.~Kovalchuk}	
\affiliation{NRC Kurchatov Institute, Moscow, Russia}
\affiliation{IC RAS, Moscow, Russia}

\date{\today}

\begin{abstract}
We present a model independent approach for the reconstruction of the atomic concentration profile in a nanoscale layered structure, as measured using the X-ray fluorescence yield modulated by an X-ray standing wave (XSW). The approach is based on the direct regularized solution of the system of linear equations that characterizes the fluorescence yield. The suggested technique was optimized for, but not limited to, the analysis of periodic layered structures where the XSW is formed under Bragg conditions. 

The developed approach was applied to the reconstruction of the atomic concentration profiles for $LaN/BN$ multilayers with 50 periods of 35~\AA \ thick layers. The object is especially difficult to analyse with traditional methods, as the estimated thickness of the interface region between the constituent materials is comparable to the individual layer thicknesses. However, using the suggested technique it was possible to reconstruct the $La$ atomic profile, showing that the $La$ atoms stay localized within the $LaN$ layers and interfaces and do not diffuse into the $BN$ layer. The atomic distributions were found with an accuracy of 1~\AA. The analysis of the $Kr$ fluorescence yield showed that Kr atoms originating from the sputter gas are trapped in both the $LaN$-on-$BN$ and the $BN$-on-$LaN$ interfaces.
\end{abstract}


\maketitle

\section{ Introduction}
The X-ray standing wave (XSW) technique~\cite{1ak, 2kk} is applied to non-destructively reconstruct the atomic profiles in crystals and in periodic~\cite{3zk, 4z, 5tsl, 6gd, 7z} or aperiodic stratified structures~\cite{3zk, 4z, 8b, 9g, 10r}. The technique is based on the measurement and analysis of the characteristic signal from specific atoms excited by the XSW formed inside a structure. The position of nodes and antinodes of the XSW formed at Bragg reflection conditions in a periodic layered structure or at total external reflection conditions in a non-periodic structure can be modified by changing the incidence angle. The angular dependent intensity of secondary emission yield from the atoms excited by the XSW is now determined by the overlap between the atomic profile and the electromagnetic field. Knowing the electromagnetic field distribution inside the structure, the atomic distribution can be reconstructed from the measured angular dependent fluorescence yield. 

In this paper we consider the XSW analysis using X-ray fluorescence. The most reliable approach to the atomic profile reconstruction is a simultaneous fit of grazing incidence X-ray reflectivity (GIXR) and XSW data having the atomic profile as a fit parameter. However, this technique is time consuming because of the large amount of fit parameters, and moreover, the outcome may be dependent on the initial model. The complicated data analysis is generally the limiting factor for the application of the XSW technique. 

Recently, a model independent approach to the reconstruction of the atomic distribution profile from XSW data was suggested by Cheng et al.~\cite{11c} and later extended by Kohli et al.~\cite{12kbf}. The work~\cite{11c} presents the Fourier transformation of measured fluorescence yields excited by the Bragg-XSW in a single crystalline sample. The Fourier transformation requires the measurement of the angular dependent fluorescence yield at different Bragg reflection orders and therefore requires a highly ordered structure. An extension of the this approach for the analysis of thin film structures with long period XSW was presented by Kohli et al.~\cite{12kbf}. This method relies on the XSW data with a large number of fluorescence yield oscillations, and imposes strict requirements on the design of the sample to be analyzed. In the current paper, we present a new approach for a model independent analysis of the XSW data that is based on the direct solution of the ill-defined system of linear equations describing the angular dependent fluorescence yield using the Tichonov regularization technique~\cite{13ta}. Similarly to V.~Kohli et al.~\cite{12kbf}, the presented analysis requires the knowledge of the electromagnetic (EM) field that can be obtained from the analysis of grazing incidence X-ray reflectivity data~\cite{14y}. 

In this paper we will use XSW to analyze the atomic concentration profiles of $La$ and pollutant ($Kr$) atoms in short period $LaN/BN$ multilayer mirrors. Such multilayers are considered as very promising reflective optical coatings for a next generation EUV lithography~\cite{15m, 16p} at 6.7~nm wavelength and their optical performance is intrinsically linked to the in-depth atomic profiles. The preliminary structural analysis of $La/B$-based multilayer stacks ($La/B$ and $La/B_4C$) shows high interface imperfections because of the intermixing between the $La$ and $B$ layers~\cite{15m}. The passivation of $La$ with $N$ ions improves the quality of multilayer mirror~\cite{17t}. It can be expected that passivation of both layers has the potential to create diffusion free multilayers because of the chemical stability of $LaN$ and $BN$. 

Especially small fluctuations in the atomic profiles pose a challenge to traditional analysis techniques such as transmission electron microscopy and X-ray reflectometry. We will show that the XSW technique has a unique capability to resolve these details. In this paper, the XSW technique was mainly applied to study the penetration of $La$ into $BN$ layers in $LaN/BN$ multilayers. Additionally the XSW technique was applied to analyze the distribution of the $Kr$ contamination inside the multilayer period.

\section{ Modeling}
\subsection{XSW data analysis}
The intensity of the fluorescence yield from atoms in a film is determined by the electromagnetic field ${\left|E(\theta,z)\right|}^2$, depending on the incidence angle~$\theta$ and distance~\textit{z} from the film surface, and the atomic distribution profile~\textbf{P} in the film. For calculations of the EM field it is necessary to divide the entire film into very thin sub-layers where the thickness of individual sub-layers is much smaller than the thickness of the layers, with each sublayer having a constant atomic concentration. In the dipole approximation the angular dependence of the fluorescence yield intensity $Y(\theta)$ is calculated as the sum over all sublayers of the products of the electromagnetic field distributions $\left|E(\theta,z_j)\right|^2$ and the concentration of fluorescent atoms in each sub-layer $P_j$, corrected for the geometrical factor $G(\theta)$ and for absorption of the fluorescence radiation: 

\begin{equation}
Y(\theta)=G(\theta){\sum_{j}P_{j}\left|E(\theta ,z_{j} )\right|^2} e^{-\mu_{f} z_{j} }. 	
\label{eq1}
\end{equation}

Here $\mu_f$ is the averaged linear absorption coefficient at the fluorescence wavelength on the exit path from the film. The geometrical factor takes into account the variation of the beam footprint with the change of the incidence angle. 

If the studied sample is a periodic multilayer structure that contains \textit{N} bi-layers with thickness $\Lambda$, the atomic profile~\textbf{P} will have~\textit{N} identical periods. Assuming a perfect periodicity, the multilayer can be presented as one ``effective'' period where the electromagnetic field distribution is the summed EM field from all periods in the multilayer. The effective EM field that excites fluorescence for the whole multilayer can now be represented as

\begin{equation}
\bar{I}_{j}(\theta)=G(\theta)\sum_{k=1}^{N}\left|E(\theta,z_{jk})\right|^{2}  e^{-\mu _{f} z_{jk} },
\label{eq2}
\end{equation}

\noindent where $z_{jk}=D\left[k-1+(j-1/2)/m\right]$, $j = [1...m]$ is the number of the sub-layer within one period, m is the number of sublayers in one period and $k = [1...N]$ is the number of the period in the multilayer. Formula~(\ref{eq1}) can then be simplified to

\begin{equation}
Y(\theta_{i})=\sum_{j=1}^{m}{P{'}}_{j} \bar{I}_{j}(\theta _{i} ).
\label{eq3}
\end{equation}

In equation~(\ref{eq3}), $P{'}$ is the atomic distribution along a period. For brevity, the apostrophe will further be omitted. 

Having measured the fluorescence angular dependency $Y_{exp}$, according to the method of least squares the unknown profile $P_j$ is found by minimizing the function

\begin{equation}
	\chi ^{2} =\frac{1}{n-m} \sum _{i}\frac{1}{\sigma _{i}^{2} } \left(Y_{\exp } \left(\theta _{i} \right)-P_{j} \bar{I}_{j} \left(\theta _{i} \right)\right)^{2},  
\label{eq4}
\end{equation}

\noindent where \textit{n} is the number of measured angular points and $\sigma_i$ the statistical error of the fluorescence yield measurements.

Generally, $\chi^2$ can be minimized using a fit procedure if there is no algebraic solution possible. However in the case presented here, the problem can be presented as a system of linear equations

\begin{equation}
{\partial\chi^2/\partial{P_j}=0}.
\label{eq5}
\end{equation}

After taking the derivative of $\chi^2$, eq.~(\ref{eq5}) is transformed into~\cite{18c}

\begin{equation}
\hat{A}\mathbf{P}=\mathbf{b},
\label{eq6}
\end{equation}

\noindent where 

\begin{equation}
\hat{A}_{jl} =\sum_{i=1}^{n}\frac{\bar{I}_{j} (\theta _{i} )\bar{I}_{l} (\theta _{i} )}{\sigma _{i}^{2} },	
\label{eq7}
\end{equation}

\begin{equation}
b_{l} =\sum _{i=1}^{n}\frac{Y_{\exp } (\theta _{i} )\bar{I}_{l} (\theta _{i} )}{\sigma _{i}^{2} },
\label{eq8}	
\end{equation}

\noindent and $j,l = [1...m]$, the number of the sublayer in a multilayer period.

The system of equations~(\ref{eq6}) can easily be solved numerically for~\textbf{P}. However, due to systematic and statistical experimental errors the reconstructed atomic depth profile may exhibit non-physical features such as negative values and strong fluctuations. This effect is the consequence of an ill-posed problem~\cite{13ta}. In order to obtain reasonable solutions, a regularization is introduced in the solving algorithm. The limitation can e.g. be the profile smoothness should be smooth. Mathematically this requirement can be introduced by adding the auxiliary term \textit{u} in the function $\chi^2$. We use the following auxiliary term: 

\begin{eqnarray}
\label{eq9}
	u=\lambda \Biggl[\sum _{j=2}^{m-1}(2P_{j} -P_{j-1} -P_{j+1} )^{2} + \cdots \nonumber \\ \cdots +(2P_{1} -P_{m} -P_{2} )^{2}+ \cdots  \\ \cdots +(2P_{m} -P_{m-1} -P_{1} )^{2} \Biggl], \nonumber
\end{eqnarray}

\noindent wherein the first term defines the smoothness of the profile amplitude within a period. The second and third terms define continuity on the upper and lower interface, respectively. The smoothening coefficient $\lambda$ determines the maximum ``allowed'' changes from $P_j$ to $P_{j+1}$ and should be selected for each separate case individually. The system of equations~(\ref{eq6}) is then transformed into the system of linear equations

\begin{equation}
	(\hat{A}+\lambda \hat{D})\mathbf{P}=\mathbf{b}.
\label{eq10}
\end{equation}

Here $\hat{D}$ is the regularization Gram matrix:

\begin{equation}
	\hat{D}=\left(
	\begin{array}{ccccc} 
	{2} & {-1} & {0} & {\cdots } & {-1} \\ 
	{-1} & {2} & {-1} & {} & {0} \\ 
	{\vdots } & {\ddots } & {\ddots } & {\ddots } & {\vdots } \\ 
	{0} & {} & {-1} & {2} & {-1} \\ 
	{-1} & {\cdots } & {0} & {-1} & {2} 
	\end{array}
	\right),
\label{eq11}
\end{equation}

The used Gram matrix is adapted for the periodical structure of multilayer mirrors. A similar type of regularization, although for non-periodic structures, has been successfully applied for the similar problem of reconstructing scatter density profiles from X-ray reflectometry~\cite{19hgs}. 



The methods and procedures described here will be applied in the experimental section of the paper. Using the described algorithms the XSW data can be analyzed without any pre-assumptions about the atomic distribution.

\subsection{Calculation of the EM field}
The electron density profile of the periodic multilayer structure was reconstructed by iterative fitting of model-based reflectivity calculations to the measured GIXR data. In the simplest case a layer in a multilayer film can be descried with 4 parameters: layer thickness, interface thickness, layer density and a material composition. The changes in the electron density profile at the interfaces are described by dividing the interface region in equal sub-layers with a thickness of less than 1~\AA\ and assuming a linear transition in the electron density between neighbouring layers. In this model each interface will still be described with one parameter: the thickness of the region of linear transition. A more complex parameterized description of the electron density profile in the interface transition regions could be used, but goes beyond the scope of this work. This description of interfaces as presented here is preferred over the standard Debye Waller or Nevot-Croce approaches when the calculation of the EM fields in the interface regions is required. 

The reflectivity calculations were performed using the Abeles matrix formalism~\cite{20a} which requires the multiplication of the characteristic matrices that describe electromagnetic wave penetration through all the layers in the sample, including the sub-layers in the interface regions. If the multilayer has good periodicity the and periods can be assumed identical, the Chebishev polynomials can be used~\cite{21bw} to calculate analytically the $\mathrm{N^{th}}$ power of the characteristic matrix for each period in the multilayer. This limits the number of matrix multiplications to the calculation of the characteristic matrix for one period only. This approach can be applied if the errors introduced in the parameter determination by aperiodicity of the multilayer are within the general uncertainty of parameter determination.

\section{ Experimental}

The GIXR data were measured using a PANalytical X'Pert PRO diffractometer using $CuK_{\alpha 1}$ radiation and a 4x$Ge$(220) asymmetrically cut monochromator. The measurements were done using a constant $2\theta$ step of 0.005 degrees and 2 seconds of exposure at each step. The XSW measurements were performed at the Hasylab E2 beamline of the DESY synchrotron radiation facility. The bending magnet radiation was monochromized for the wavelength of 0.71~\AA. The fluorescence spectra were measured for 5 seconds per angular step using a Roentec energy dispersive detector. Angular scans were repeated in a fixed range of angles around the first Bragg reflection peak and the signals were summed until the statistical error of the accumulated integral fluorescence yield was better than 1\%. 

The 50 period $LaN/BN$ multilayer structure was deposited using DC magnetron sputtering of $La$ and $B_4C$ targets using $Kr$ as a sputter gas. This number of period was selected as a compromise between the number of periods required for the formation of the contrast X-ray standing wave, the excitation of sufficiently intense fluorescence radiation and the stability of the deposition process. The passivation of $La$ and $B_4C$ layers was done using low energy $N$-ion treatment of the deposited layers~\cite{22l}. XPS measurements were used to optimize the parameters of metrication for the deposition of $LaN$ and $BN$ films (optimization is not presented here). The ratio of $La$ layer to period thickness was selected 0.4 for optimal optical performance.  

\section{Results}
\subsection{Electron density profile reconstruction }

The electron density profile was reconstructed from the fit of the GIXR data in order to calculate the EM field for XSW analysis. The analysis of GIXR data was performed in two steps: first the fit was performed with a model that describes a 50 times repeated structure, the multilayer period, varying the period structure parameters (layer and interface thicknesses and densities), and assuming no period variation along the stack. Separately, a second fit was performed where the parameters of all periods were varied individually. From the first fit the average period structural parameters and their statistical errors were obtained. From the second fit the individual parameters of all layers were obtained and their deviation from the average was determined. Comparing both the averaged and individual parameter sets allows checking the applicability of the model with identical periods for the XSW analysis. 

\begin{figure}[b]
\includegraphics[width=1\linewidth]{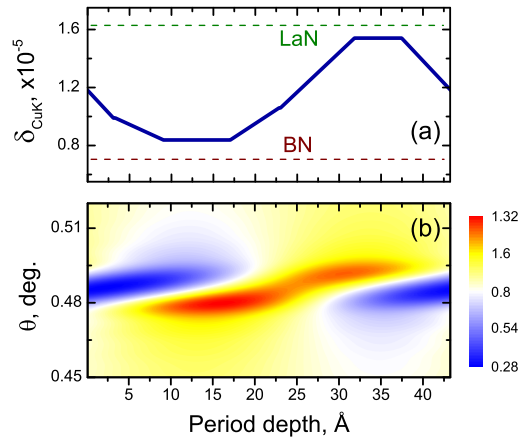}
\caption{Reconstructed electron density profile in the $LaN/BN$ multilayer (a) and EM field distribution in the vicinity of the 1-st Bragg peak (b) calculated for the wavelength of 0.71~\AA \ used in the fluorescence yield measurements.}
\label{fig1}
\end{figure}

The average values were obtained for the model containing 49 identical periods. The parameters of the top layers were fitted separately assuming that contact with ambient gases possibly changes the top layer structure. The best model representation of the electron density profile is shown in Fig.~\ref{fig1}a. As depicted in Fig.~\ref{fig1}a, each interface consists of two linear segments with different slopes, indicating that one simple linear transition is not good enough. For each parameter the fit errors were estimated. The parameter values and their errors are shown in table~\ref{table1} in the column ``Average''. 

\begin{figure}
\includegraphics[width=1\linewidth]{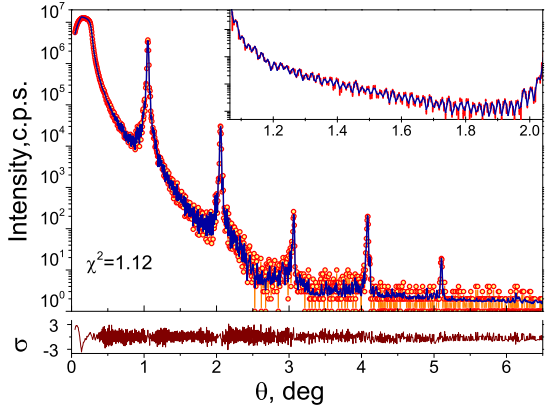}
\caption{Experimental and simulated GIXR data for the model with individually defined periods. The insert shows the measured and simulated Kiessig fringes between first and second order Bragg peaks.}
\label{fig2}
\end{figure}

In the second fit we allowed all 50 periods to have individual layer parameter. This ``individual layers'' fit improved mostly because of the better match of the Kiessig fringes (Fig.~\ref{fig2}). Analysing the distribution of individual layer parameters, their standard deviation from their average values is determined and presented in table 1 in the column ``Deviation''. The obtained period thickness of the multilayer determined from both fits was $\Lambda=43.4$~\AA \ with a standard deviation of the many-period fit $\mathrm{s_D~=~0.24}$~\AA, close to the error in the period determination from the fit of the model with identical periods and small compared to the actual value. This confirms that the model with identical periods describes the density profile in the full stack well and may be used for XSW data analysis. The EM field was finally calculated using the Abeles matrix formalism for the reconstructed averaged multilayer profile and summed along the \textit{z} direction over all periods according to eq.~(\ref{eq2}). Figure~\ref{fig1}b shows the averaged EM field visualizing positions of the nodes and antinodes of the XSW within a period. 

From the GIXR profile it is observed that approximately 70\% of the period thickness is in the interface state (referring to the two gradients in between the $LaN$ and $BN$). This would suggest that there is significant intermixing in the multilayer. 

\begin{table*}
\caption{$LaN/BN$ model parameters reconstructed from GIXR. For each parameter we show averages over all the periods and a standard deviation of the values for all 50 periods.}
\label{table1}
\renewcommand{\arraystretch}{1.8} 
\begin{tabular}{|c|c|c|c|c|c|c|}

\hline
	\multicolumn{1}{|c|}{ } 
	& \multicolumn{2}{c|}{Layer} 
	& \multicolumn{2}{c|}{The interface transition} 
	& \multicolumn{2}{c|}{Density,} \\
	
	\multicolumn{1}{|c|}{ } 
	& \multicolumn{2}{c|}{thickness,  \AA} 
	& \multicolumn{2}{c|}{layer thickness,  \AA} 
	& \multicolumn{2}{c|}{g/cm$^3$} \\

\cline{2-7} 
	\multicolumn{1}{|c|}{ } 
	& \multicolumn{1}{c|} {Average} 
	& \multicolumn{1}{c|} {Deviation} 
	& \multicolumn{1}{c|} {Average}  
	& \multicolumn{1}{c|} {Deviation} 
	& \multicolumn{1}{c|} {Average}  
	& \multicolumn{1}{c|}{Deviation} \\ 

\hline 
	\multicolumn{1}{|c|}{$BN$} 
	& \multicolumn{1}{c|}{7.95$\pm$0.16} 
	& \multicolumn{1}{c|}{0.19} 
	& \multicolumn{1}{c|}{14.9$\pm$0.2} 
	& \multicolumn{1}{c|}{0.2} 
	& \multicolumn{1}{c|}{2.7 $\pm$ 0.05} 
	& \multicolumn{1}{c|}{0.1} \\ 

\hline 
	\multicolumn{1}{|c|}{$LaN$} 
	& \multicolumn{1}{c|}{5.63$\pm$0.08} 
	& \multicolumn{1}{c|}{0.21} 
	& \multicolumn{1}{c|}{14.8$\pm$0.2} 
	& \multicolumn{1}{c|}{0.3} 
	& \multicolumn{1}{c|}{5.8 $\pm$ 0.1} 
	& \multicolumn{1}{c|}{0.13} \\ 
	
\hline 

 \end{tabular}
\end{table*}

\subsection{Atomic distribution profile reconstruction}

\begin{figure}[b]
\includegraphics[width=1\linewidth]{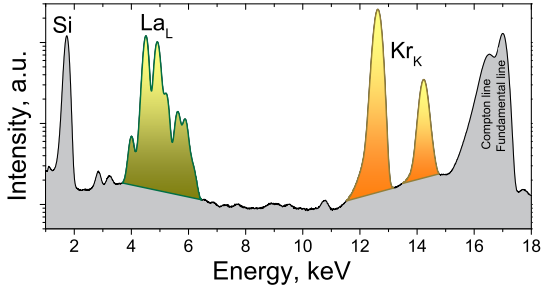}
\caption{Typical X-ray fluorescence spectrum measured from the $LaN/BN$ multilayer.}
\label{fig3}
\end{figure}

A typical measured fluorescence spectrum is presented in Fig.~\ref{fig3}. Additional to the expected $La~L$ fluorescence yield, fluorescence from $Si$ (substrate) and $Kr$ was detected. $Kr$ was used as the magnetron sputtering gas, and its presence indicates trapping of $Kr$ in the multilayer. The small doublet at 3~keV originates from the $Ar~K_\alpha$ and $K_\beta$ lines. The intensity of this signal corresponds to the intensity of $Ar$ from the ambient environment. For XSW analysis, the background corrected integral intensity of $La~L$ and $Kr~K$ fluorescence radiation were determined at a range of angles of incidence around the first order Bragg reflection. 

\begin{figure}
\includegraphics[width=1\linewidth]{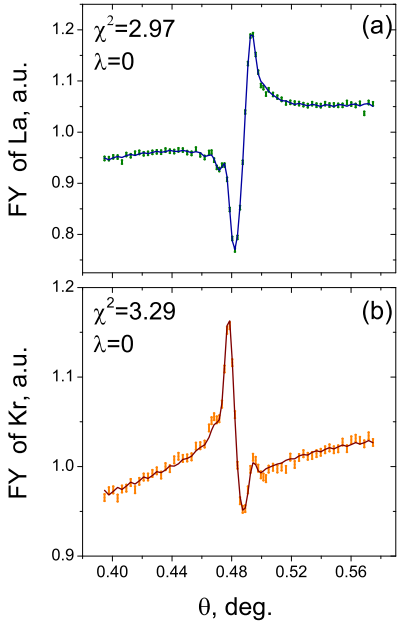}
\caption{Experimental (symbols) and simulated (solid lines) X-ray fluorescence yields (FY) for $La$ (a) and for $Kr$ (b) in the region of the first Bragg peak region.}
\label{fig4}
\end{figure}

\begin{figure}
\includegraphics[width=1\linewidth]{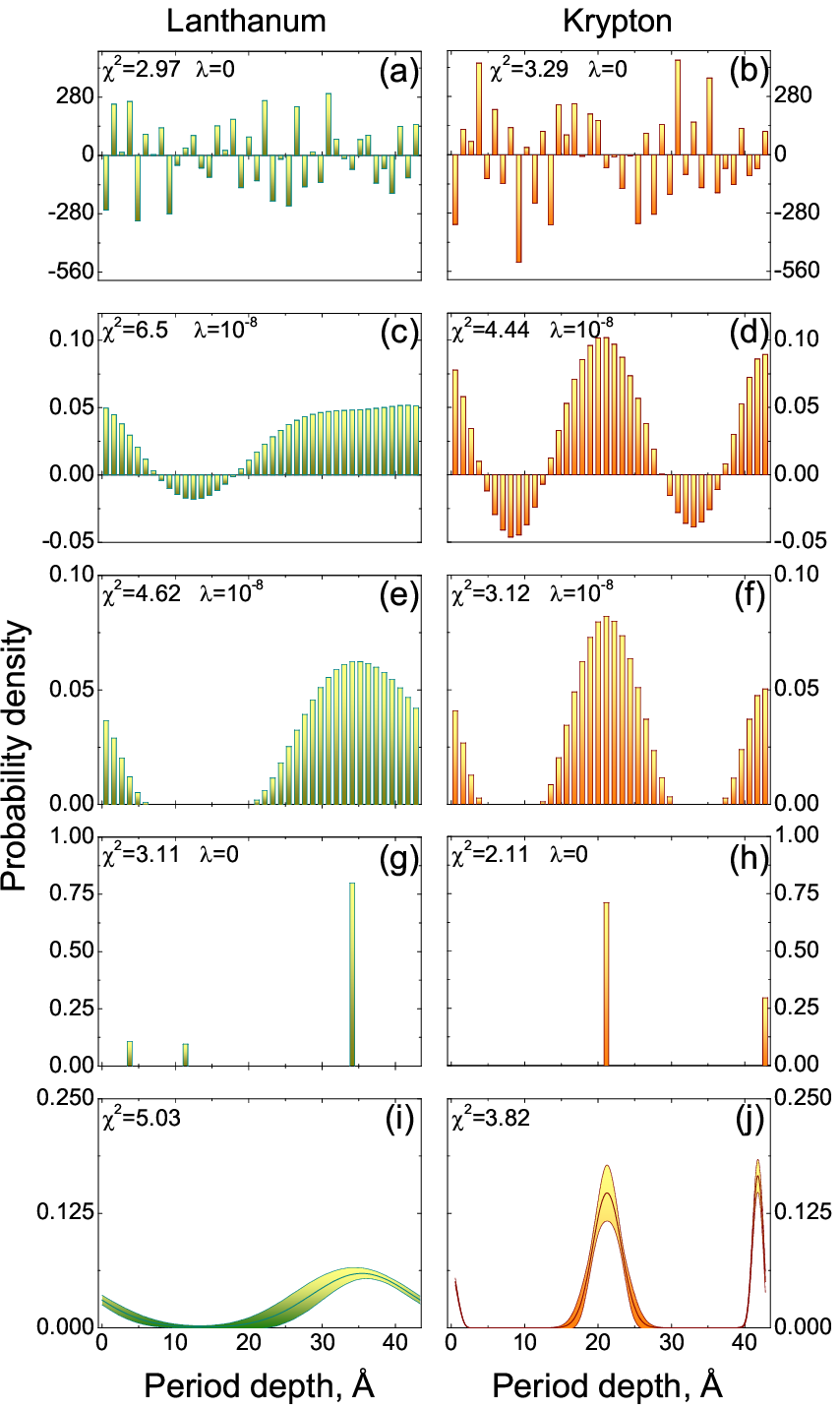}
\caption{Reconstructed atomic profiles for $La$ and $Kr$ using no profile limitations (a and b), profile smoothening (c and d), and smoothening with iterative removal of sublayers with negative values (e and f). Also shown are non-smoothened profiles with iterative removal of sublayers with negative values (g and h) and profiles obtained using Gaussian atomic distributions (i and j).}
\label{fig5}
\end{figure}

The angular dependencies of the $Kr$ and $La$ fluorescence yields calculated based on the direct solution of the system of linear equations~(\ref{eq6}), using EM fields reconstructed from the GIXR measurements, are presented in Fig.~\ref{fig4}. The corresponding reconstructed $La$ and $Kr$ atomic density profiles are presented in Figures~\ref{fig5}a and~\ref{fig5}b. Note that all atomic density profiles presented in Fig.~\ref{fig5} are normalized such that the integral of the profile is unity, corresponding to the probability density of the atom distribution in the period of the multilayer structure. Although a very good agreement between simulations and experiments is observed in Fig.~\ref{fig4}, the reconstructed atomic profiles are clearly non physical when no limitations are introduced in the profile, as can be observed from the negative probabilities in Figs.~\ref{fig5}a and~\ref{fig5}b. Note that even for a non physical solution the fit goodness for $La$ ($\chi^2$=2.97) and for $Kr$ ($\chi^2$=3.29) are not equal to unity. For other solutions we will not present the calculated curves but will indicate the values for the obtained $\chi^2$ that need to be compared with the values obtained for a non regularized solution. 

To resolve the ill-posed problem, limitations to atomic profiles were introduced according to eq.~(\ref{eq10}). For smoothening of the profile, the coefficient $\lambda=10^{-8}$ was used for both element distributions. This value for $\lambda $ was selected to provide the best fit of the measured to calculated XSW data. The fit quality for smoothed profiles is just slightly worse than for not smoothed: $\chi^2$=6.5 for $La$ and $\chi^2$=4.4 for $Kr$. The resulting atomic profiles are presented in Figs.~\ref{fig5}c and~\ref{fig5}d. 

Note that the smooth profile is still physically impossible because of the locally negative density values for both materials. The smoothening of layers can force the profile to have negative atomic concentrations to enable artificially smoothed transition between the regions with and without atoms where there are natural profile distributions. The layers with negative atomic concentrations should then be cancelled by sequential "removal" of the individual equations that correspond to the sub-layers with the largest negative concentrations, followed by searching for a new solution of a lower-rank system of equations. After the sequential removal of negative probability densities and solving of the reduced system of equations~(\ref{eq10}), the final profiles were found and are shown in Figs.~\ref{fig5}e and~\ref{fig5}f. The fit goodness of the final fit ($\chi^2$=4.6 for $La$ and $\chi^2$=3.1 for $Kr$) is actually better than that for the initial smoothed profile that allowed negative probability densities because of the reduced degrees of freedom during the calculations of the $\chi^2$ function. 

To determine the accuracy and stability of the reconstructed profiles, additional analysis was performed. Unfortunately the suggested approach (eq.~\ref{eq10}) does not allow evaluation of errors in the reconstructed profiles because the shape of the reconstructed profile is dependent on the smoothening parameter $\lambda$ which influences errors. To estimate errors and the stability of the profile determination, the XSW data can be fitted using a Gaussian shape of the atomic concentration distribution. Errors can be derived from this fit. The fit requires the input of initial parameters for the Gaussian model where the exact positions of a distribution center are the most important model parameters. These positions can be found by the iterative removal of sub-layers with negative values in atomic concentrations obtained from the solution of the non regularized system of linear equations~(\ref{eq6}).

Starting from the non-smoothened profiles (Figs.~\ref{fig5}a and~\ref{fig5}b), sub-layers with the highest negative value were iteratively removed until no negative values remained, resulting in the profiles shown in Figs.~\ref{fig5}g and~\ref{fig5}h. The obtained profiles correspond to the peak positions of $La$ distribution profiles. The thus determined profiles were further resolved by the fit of model based calculations of fluorescence yield to measured XSW data, assuming that concentration profiles follow a Gaussian distribution form with peak center position and peak width as fit parameters. The center positions of the initial Gauss profiles were obtained from Figs.~\ref{fig5}g and~\ref{fig5}h. Because of the relative importance of the single peak at $z=35$~\AA, a single Gaussian distribution at this position was used for the analysis of the $La$ profile. The result of the reconstruction is presented in Fig.~\ref{fig5}i. The fluorescence yield of $Kr$ was simulated with two separate Gaussians located at the positions obtained from Fig.~\ref{fig5}h. Comparing the $La$ profiles from Figs.~\ref{fig5}e and~\ref{fig5}h we can conclude that all exhibit the same location of $La$ within the period, within a 1~\AA \ accuracy. The same conclusion can be drawn for $Kr$.

\section{Discussion}
The model-independent technique for XSW data analysis presented here expands the series of model independent approaches presented in works Cheng\cite{11c} and Kohli\cite{12kbf} and completes the set of model independent approaches for all types of XSW techniques: Bragg-XSW for single crystals, long period XSW for layered structures and Bragg XSW for periodic multilayer structures.

The benefits of the current approach is that for the reconstruction of atomic profiles by direct solution of Eq.~(\ref{eq10}) the XSW data can be measured only for one order of Bragg reflection. The disadvantage is that regularization procedure forces the profile to be smoothed and inaccurate selection of the smoothening coefficient \textit{l} may force an artificially smoothened profile. Taking into account that in Eq.~(\ref{eq6}) the fluorescence yield is excited by the general EM field shape, the current approach can be extended to the long-period XSW technique. However  the modification of the regularization technique might be required there. 

The analysis of errors in the profiles reconstructed using the XSW technique shows that the position of the maximum in the atomic distribution profile can be determined with an accuracy of 1~\AA. The analysis of errors was performed assuming that the EM field does not change with the variation of the atomic profiles, and suggests that the derived error is slightly underestimated. However if all experimental artifacts connected to the beam spectral and geometrical resolution and goniometric uncertainties are taken into account, the reconstructed positions of atom localization are reliable. We should note that because of the shape of the EM field, an increase of the atomic distribution width will lead to a decrease in the accuracy of the shape of the distribution, reducing the accuracy of the profile width determination. We also note that the XSW technique yields the averaged over all periods profile, and a strong aperiodicity in the sample will therefore be misinterpreted as a blurring of the atomic distribution. The GIXR technique applied here for the EM field reconstruction is sensitive to the periodicity of the multilayer and will allow estimation of period fluctuations before the XSW analisys is performed. 

\begin{figure}
\includegraphics[width=1\linewidth]{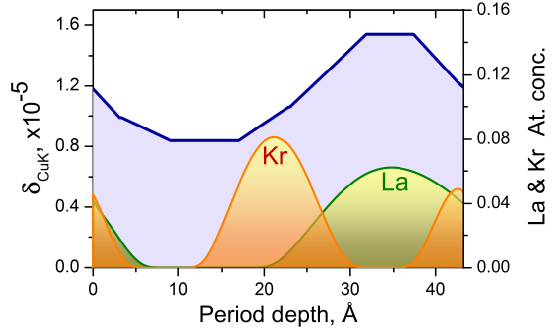}
\caption{Comparison of obtained electron density and atomic profiles.}
\label{fig6}
\end{figure}

Figure~\ref{fig6} shows a comparison between the optical contrast profile obtained from GIXR and the atomic distribution profile obtained from the XSW analysis. As discussed in the previous section, the atomic distribution is presented in terms of a normalized probability density per element. In absolute value, the amount of $Kr$ is approximately one order of magnitude less than that of~$La$. 
In apparent contrast to the suggestion made based on the GIXR analysis about the intermixing between $LaN$ and $BN$ layers, the XSW analysis shows that $La$ is well localized and the width of the $La$ distribution corresponds to 40\% of the period thickness, as expected from the deposition design. The absence of $La$ atoms in the $B$ layer indicates that the metrication of both layers helps to prevent the $La-B$ intermixing in the stack. 

The unique result of the XSW analysis is a non destructive analysis of the impurity distribution. A small residue of the sputtering gas $Kr$ could be detected and appears localized in the interface regions. Apparently, $Kr$ ions from the magnetron plasma (with energies~300~eV during $La$ sputtering and~600~eV during $B_4C$ sputtering) are capable to penetrate through the already deposited layers and be trapped in the interfaces. The result is also important as $Kr$ has a high absorption for the 6.7~nm wavelength and its presence will reduce the reflectance in the envisioned application. 

\section{Conclusions}
As demonstrated in this paper, the analysis of X-ray standing wave data from periodic multilayer structures based on the solution of linear equations describing the fluorescence yield, allows a fast and model independent reconstruction of atomic profiles. The approach was applied to Bragg-XSW data from a $LaN/BN$ multilayer structure. The obtained profile of the $La$ atoms in the structure showed an accurate, 1~\AA \ localization of $La$ atoms within the $LaN$ layer and the absence of La atoms in the $BN$ layer. Additionally, the contamination of the multilayer by $Kr$ atoms, trapped during the magnetron deposition process, was revealed. It was found that these $Kr$ atoms are distributed inside the interface regions. The sensitivity of the XSW technique to such atoms remains high, even though their presence does not change the electron density profile. This demonstrates that XSW is capable to provide non-destructive depth resolved elemental composition with sub-nm accuracy.

\begin{acknowledgments}
The part of this work belongs to the project ``Multilayer Optics for Lithography Beyond the Extreme Ultraviolet Wavelength Range'' carried out with support of the Dutch Technology Foundation (STW) in the frame of the Thin Film Nanomanufacturing programme. This work is also part of the research programme ``Controlling photon and plasma induced processes at EUV optical surfaces (CP3E)'' of the ``Stichting voor Fundamenteel Onderzoek der Materie (FOM)'' with financial support from the ``Nederlandse Organisatie voor Wetenschappelijk Onderzoek (NWO)'', Carl Zeiss SMT, ASML, and the AgentschapNL through the EXEPT program.

The part of this work was supported by a grant for Russian young scientists and leading scientific schools NSH - 5837.2012.2 ``Physics of coherent interaction of X-ray and synchrotron radiation with matter, the development of high-resolution phase-sensitive X-ray techniques for structural diagnostics of crystalline and nanomaterials'' and Grant of Russian Foundation for Basic Research number 12-02-12045-ofi-m.
\end{acknowledgments}

\bibliography{XSW_Yakunin}

\end{document}